\documentclass[twoside,twocolumn,english,aps,prx]{revtex4-1}

	\usepackage[usenames,dvipsnames]{xcolor}
	\usepackage{amsmath}
	\usepackage{amsfonts}
	\usepackage{amssymb}
	\usepackage[colorlinks=true,citecolor=blue,linkcolor=red]{hyperref}
	\usepackage{graphicx}
	\usepackage{bbold}					
	\usepackage[makeroom]{cancel}		
	\usepackage{multirow}				
	\usepackage[normalem]{ulem}        
	\usepackage{array}
	\usepackage{comment}
	\usepackage{pdfpages}
	\makeatletter
	\AtBeginDocument{\let\LS@rot\@undefined}
	\makeatother
	
	\newcolumntype{x}[1]{>{\centering\let\newline\\\arraybackslash\hspace{0pt}}p{#1}}
	\renewcommand{\Re}{\operatorname{Re}}

	\DeclareMathAlphabet{\mathbbold}{U}{bbold}{m}{n}

	\newcounter{subeqn} %
	\makeatletter
	\@addtoreset{subeqn}{equation}
	\makeatother

\definecolor{TB}{rgb}{0,0,0} 

\begin{document}
\title{Quantum Error Correction Alleviates Detection-Induced Coherent Errors}

\author{Qinghong Yang$^{1}$}
\author{Dong E. Liu$^{1,2,3}$}\email{Corresponding to: dongeliu@mail.tsinghua.edu.cn}

\affiliation{$^{1}$State Key Laboratory of Low Dimensional Quantum Physics, Department of Physics, Tsinghua University, Beijing, 100084, China}
\affiliation{$^{2}$Beijing Academy of Quantum Information Sciences, Beijing 100193, China}
\affiliation{$^{3}$Frontier Science Center for Quantum Information, Beijing 100184, China}

\date{\today}


\begin{abstract}
We study the performance of quantum error correction codes(QECCs) under the {\em detection-induced coherent error} due to the imperfectness of practical implementations of stabilizer measurements, after running a quantum circuit. Considering the most promising surface code, we find that the detection-induced coherent error will result in undetected error terms, which will accumulate and evolve into logical errors. However, we show that such errors will be alleviated by increasing the code size, akin to eliminating other types of errors discussed previously. We also find that with detection-induced coherent errors, the exact surface code becomes an approximate QECC. 
\end{abstract}

\maketitle

\section{Introduction}
The advent of programmable quantum computers~\cite{arute,Arute-Google2020Science,Gong-USTC2021-Science,Pino-TrapIon2021-Science,wu2021strong,ryananderson2021} using noisy intermediate-scale quantum (NISQ)~\cite{Preskill2018NISQ} processors has recently shown strong evidence of quantum ``supremacy" or quantum advantage~\cite{arute,Zhong-USTC2020-Science,wu2021strong}. Those quantum hardware achievements along with the software development~\cite{QuantumSoftware-Nature2017-Review} pave the way for the potential fault-tolerant universal quantum computation. It is believed that the next milestone will be the experimental demonstrations of fault-tolerant quantum error corrections (QEC)~\cite{shor,stean,Calderbank96}, and  some important progresses~\cite{linke,andersen,mcewen,googleai} have been reported recently. 

In the past twenty years, people have invented many QECCs, such as the $9$-qubit Shor code~\cite{shor}, the $7$-qubit Steane code~\cite{stean}, and the most promising surface code~\cite{kitaev,dennis,fowler}. A remarkable result of QEC is the threshold theorem, which states that if the physical error rate is below a critical value, QECCs can suppress the logical error rate to an arbitrarily low level~\cite{knill2,aharonov2,aliferis,nielsen}. The threshold theorem is proved based on stochastic error models~\cite{dennis,fowler,bombin,stephens,tomita,tuckett}, and the effectiveness of QECCs is also confirmed later from discussions of the correlated noise model~\cite{aharonov,novais,novais2,benaroya,chubb,staudt,chao} and the coherent error model~\cite{barnes,beale,bravyi,ehuang,cai,ouyang}, assuming perfect syndrome measurements. However, due to the imperfectness of experimental implementations of quantum gates, syndrome measurements can not be perfect and this will also has some influences on QECCs. Previously, people have partially considered the faulty measurement problem, and treat it using the stochastic error model. In this case, one can repeat several rounds of syndrome measurements, roughly of order $d$ times(where $d$ is the code size of the surface code)~\cite{dennis,Shigeo}, to ensure fault-tolerance. 

\begin{figure}
    \centering
    \includegraphics[height=6.6cm]{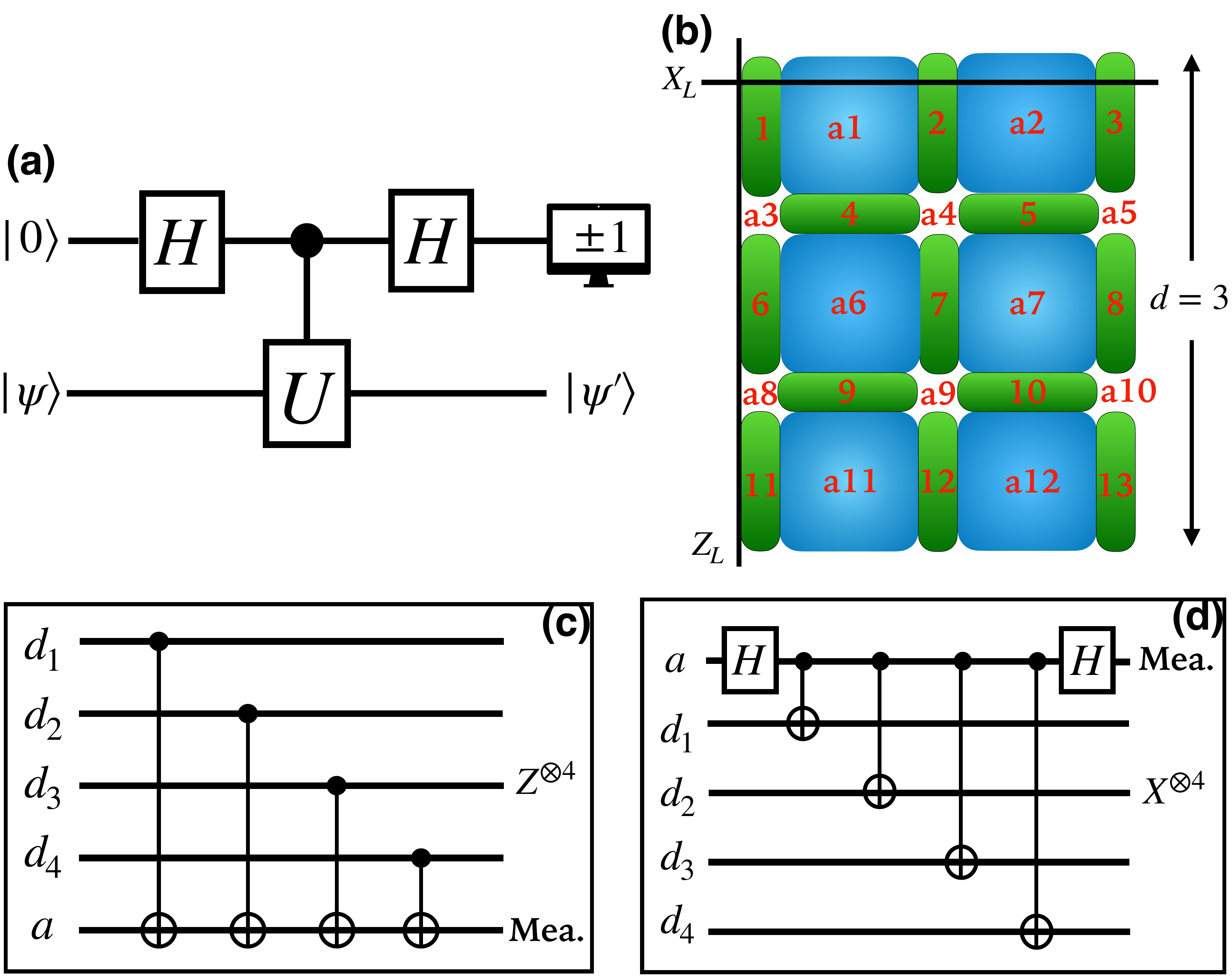}
    \caption{Illustrations of the controlled-$U$ circuit and the surface code. (a) The controlled-$U$ circuit for measuring operator $U$. (b) The $d=3$ surface code with $X_L=X_1X_2X_3$ and $Z_L=Z_1Z_6Z_{11}$ being logical operators. $i(i=1,\cdots,13)$ stands for data qubits, while $a1,a2,a6,a7,a11,a12$ are ancillas for plaquette operators ($Z^{\otimes 4}$) and $a3,a4,a5,a8,a9,a10$ are ancillas for site operators ($X^{\otimes 4}$).  (c) and (d) Circuits for measuring $Z^{\otimes 4}$ and $X^{\otimes 4}$ stabilizers of the surface code, respectively. $d_i$ and $a$ stand for data qubits and ancillary qubits, respectively.}
    \label{fig:code_illustration}
\end{figure}

Here, we focus on another type of stabilizer measurement error, dubbed as {\em detection-induced coherent error}. For a measurement of a stabilizer $U$, one can implement it using a controlled-$U$ circuit, see Fig.~\ref{fig:code_illustration}(a). However, the practical implementations may not be accurate, and the actual operation generally measures another operator $V\equiv U-\delta U$, where $\delta U$ is the deviation from the ideal one. Note that for the ion trap quantum computing  architecture, in order to combat such coherent errors during stabilizer measurements, Ref.~\cite{debroy}  introduces an extra technique, which works efficiently for the same magnitude of errors in implementations of each stabilizer. Since the connectivity of other quantum computing architectures are not as good as the trapped ion architecture, and coherent errors in each stabilizer can not perfectly be of the same order of magnitude, in reality, such coherent errors will always exist. Thus, here comes a natural question: with the existence of $\delta U$, can the final state after the QEC procedure be used for quantum computation, or in other words, can the computation using this state reach the accuracy we want? 

In this paper, through the most promising surface code(see Fig.~\ref{fig:code_illustration}(b)), we show that the detection-induced coherent error coming from stabilizer measurements of QECCs will result in accumulated logical errors, but those errors can be alleviated by QECCs themselves without extra techniques. Our discussions of the surface code are mainly concentrated on the superconducting qubit architecture \cite{arute,wu2021strong}. In addition, we also find that under the detection-induced coherent error, the exact surface code becomes an approximate QECC, thus our results imply that some approximate QECCs may also be possible for fault-tolerant quantum computation.

\section{Preliminary of detection-induced coherent errors}

Initially, the state of data qubits is supposed to be the eigenstate of all stabilizers with eigenvalue $+1$, denoted as $|\psi\rangle$. After implementing the quantum circuit shown in Fig.~\ref{fig:code_illustration}(a), we have 
\begin{equation}
    |0\rangle|\psi\rangle\rightarrow |\Psi_1\rangle\equiv \frac{1}{2}|0\rangle(|\psi\rangle+V_1|\psi\rangle)+\frac{1}{2}|1\rangle(|\psi\rangle-V_1|\psi\rangle),
\end{equation}
where $V_1=U_1-\delta U_1$ is the first stabilizer operator to be measured. Note that if there is no deviation, i.e., $V_1=U_1$, we have $|\Psi_1\rangle=|0\rangle|\psi\rangle$, as expected. For simplicity, we suppose that $V_1$ is still unitary, but not Hermitian. Since now $|\psi\rangle-V_1|\psi\rangle\neq 0$,  after measuring the ancillary qubit, one can get $-1$ with probability $p(|1-\rangle)=1/2\Re\langle\psi|\delta U_1|\psi\rangle$, 
and $+1$ with probability  $p(|1+\rangle)=1-p(|1-\rangle)$, where $|1\pm\rangle\propto|\psi\rangle\pm V_1|\psi\rangle)$, up to a normalized factor.
We'd like to know what will result from $\delta U$ in consecutive measurements, thus we consider another stabilizer measurement. Practically, it will become $V_2\equiv U_2-\delta U_2$. Suppose after the measurement of $U_1$, the state of data qubits collapses to $|1+\rangle$ (it is more likely for data qubits to collapse into this state), then the $V_2$ measurement will lead to 
\begin{equation}
    |0\rangle|1+\rangle\rightarrow\frac{1}{2}|0\rangle(|1+\rangle+V_2|1+\rangle)+\frac{1}{2}|1\rangle(|1+\rangle-V_2|1+\rangle).
\end{equation}
Letting $|2\pm\rangle\propto|1+\rangle\pm V_2|1+\rangle$, up to a normalized factor.
Considering the fidelity $F(| n\pm\rangle)=|\langle \psi|n\pm\rangle|$, we have 
\begin{equation}
    \Delta F^2=[F(|1+\rangle)]^2-[F(|2+\rangle)]^2\sim\frac{2a(a+1)}{(a-2)(5a-4)}>0
\end{equation}
for small $a$, where we have assumed that $\Re\langle\psi|\delta U_1|\psi\rangle=\Re\langle\psi|\delta U_2|\psi\rangle=a$ for the same status of $U_1$ and $U_2$. This fact shows that after two controlled-$V$ gates, the state becomes less accurate, even though syndromes show no error. Thus, one can conclude that with the proceeding of syndrome measurements, the state will become less and less accurate. We now have a concrete discussion using the surface code.

\section{Detection-Induced Coherent Errors for Surface Codes}

\subsection{Introduction to Surface Codes}\label{sub:sc}
The surface code is an example of the stabilizer codes \cite{nielsen}, and due to its locality and the high threshold \cite{dennis,fowler}, it is believed to be one of the most promising quantum error correction code for the fault-tolerant quantum computation. We briefly review the idea of the surface code~\cite{kitaev,dennis,fowler} in this subsection.

As shown in Fig. \ref{fig:code_illustration}(b), the surface code is implemented on a 2D array of physical qubits. Those physical qubits can be classified into two classes: data qubits ($1,\cdots,13$) residing on those round rectangular edges in Fig. \ref{fig:code_illustration}(b) and ancillary qubits ($a_1,\cdots,a_{12}$) residing on centers of vertices and plaquettes in Fig. \ref{fig:code_illustration}(b).  Data qubits are used to encoded quantum states, while ancillary qubits are used to detect the information, known as the error syndrome, of errors occurring on the encoded quantum state. For surface codes, we use two types of stabilizers to detect errors: $X$-stabilizers ($X^{\otimes 4}$) for detecting $Z$ errors and $Z$-stabilizers ($Z^{\otimes 4}$) for detecting $X$ errors. $X$-stabilizers appear as vertices on the 2D array, such as $X_2X_4X_7X_5$ and $X_1X_6X_4$ in Fig. \ref{fig:code_illustration}(b). If one $X$ error appears in data qubit $7$, $X_2X_4X_7X_5$ and $X_7X_9X_{12}X_{10}$ will detect it. This information is stored in ancillary qubits $a_4$ and $a_9$ through changing the states of $a_4$ and $a_9$ from $|0\rangle$ to $|1\rangle$. $Z$-stabilizers appear as plaquettes on the 2D array, such as $Z_4Z_6Z_9Z_7$ and $Z_1Z_4Z_2$. Information of $X$ errors from implementing $Z$-stabilizers is encoded on those ancillary qubits residing on the center of each plaquette. Due to the appearances of those stabilizers on the 2D array, in the following, we will call $X$-stabilizers and $Z$-stabilizers as site operators and plaquette operators, respectively. 

After implementing all stabilizers to detect errors, one should measure those ancillary qubits to obtain the error syndrome, and then use the encoding algorithms, such as the minimum-weight perfect-matching algorithm \cite{fowler}, to decode the error syndrome. The error chain can be fixed after decoding. To realize the error correction procedure, one just need to act this error chain back on the quantum state. There will be two type of errors that can not be detected: the logical $X$ error and the logical $Z$ error. Note that these two logical errors are nothing but logical $X$ and $Z$ operators for the surface code, see $X_L$ and $Z_L$ in Fig. \ref{fig:code_illustration}(b). The reason that these logical errors can not be detected is because $X_L$ and $Z_L$ commute with all stabilizers. Note that error chains differing up to some stabilizers are equivalent and can be regarded as the same error. For a complete and thorough introduction to surface codes, we refer interested readers to Ref. \cite{fowler}.


\subsection{Imperfect CNOT and Modified Stabilizer Measurements}

From the discussion in Subsec. \ref{sub:sc}, we know that for a surface code, in order to detect errors, we need to implement  two type of stabilizer operators --- $Z^{\otimes4}$ and $X^{\otimes4}$. Experimentally, each stabilizer can be realized through four $\operatorname{CNOT}$ gates (see Fig.~\ref{fig:code_illustration}(c) and \ref{fig:code_illustration}(d)), which can be constructed from two single-qubit Hadamard gates and a two-qubit controlled-Z gate denoted as $\Phi_{ct}$ (c and t are abbreviations for controlled and target qubits, respectively), that is $\operatorname{CNOT}_{ct}=\operatorname{H}_t\Phi_{ct}\operatorname{H}_t$. This strategy is widely used in superconducting quantum computation~\cite{krantz}. The controlled-Z gate can be realized through two-body interactions, which produce a time evolution operator $U(t)=\operatorname{diag}(1,1,1,\operatorname{exp}(-iJt/\hbar))$; and when $Jt_0/\hbar=\pi$, $U(t)$ reduces to the ideal contolled-Z gate $\Phi_{ct}$~\cite{krantz}. However, the fluctuation of the interaction constant $J$ (or the time interval) will produce an imperfect controlled-Z gate and then an imperfect $\operatorname{CNOT}$ gate. The actual evolution operator can be written as $U^{\prime}(t_0)=\Phi_{ct}^{\kappa}\Phi_{ct}$, where $\kappa=\delta J/J$ is the {\em imperfect rate} and $\delta J$ is the fluctuation of $J$. We assume that the fluctuation is constant (see 
App. \ref{sec:IR&MGF} for its connection with the experimentally measurable quantity). We assume the single-qubit gate is much better than the two-qubit gate~\cite{arute} and the imperfectness of $\operatorname{CNOT}$ gates only results from the imperfect controlled-Z gate, that is $\operatorname{CNOT}^{\prime}=\operatorname{H}_t\Phi_{ct}^{\kappa}\Phi_{ct}\operatorname{H}_t$. Note that the imperfect controlled-$Z$ gate was also considered in Ref.~\cite{brien,huang} for different motivations. 

Using the above error model, we can derive the modified measurement operators of stabilizers. We assume the surface code is prepared in $|\Psi\rangle$, where ancillary qubits are all in $|0\rangle$. Then, the measurements of plaquette operators (see Fig.~\ref{fig:code_illustration}(c)) can be expressed as 
\begin{equation}
\begin{split}
    &\quad\operatorname{CNOT}^{\prime}_{d_4a}\operatorname{CNOT}^{\prime}_{d_3a}\operatorname{CNOT}^{\prime}_{d_2a}\operatorname{CNOT}^{\prime}_{d_1a}|\Psi\rangle\\
    &=\left(\operatorname{H}_a\Phi_{d_4a}^{\kappa}\Phi_{d_3a}^{\kappa}\Phi_{d_2a}^{\kappa}\Phi_{d_1a}^{\kappa}\operatorname{H}_a\right)\\
    &\quad\times\operatorname{CNOT}_{d_4a}\operatorname{CNOT}_{d_3a}\operatorname{CNOT}_{d_2a}\operatorname{CNOT}_{d_1a}|\Psi\rangle,
\end{split}
\end{equation}
which means the imperfect $\operatorname{CNOT}$ introduces an extra term for measuring a plaquette operator. Keeping terms up to the first order of $\kappa$, the extra term yields 
\begin{equation}
\begin{split}
    &\quad\operatorname{H}_a\Phi_{d_4a}^{\kappa}\Phi_{d_3a}^{\kappa}\Phi_{d_2a}^{\kappa}\Phi_{d_1a}^{\kappa}\operatorname{H}_a\\
    &\sim\quad\left[(1-i\pi\kappa)\mathbf{I}^{\otimes 4}+\frac{i\pi\kappa}{4}(Z_{d_1}+Z_{d_2}+Z_{d_3}+Z_{d_4})\right]\mathbf{I}_a\\
    &\quad-\left[(0-i\pi\kappa)\mathbf{I}^{\otimes 4}+\frac{i\pi\kappa}{4}(Z_{d_1}+Z_{d_2}+Z_{d_3}+Z_{d_4})\right]X_a,
\end{split}
\end{equation}
which shows that for plaquette operator measurements, imprfect $\operatorname{CNOT}$ gates will introduce coherent $Z$ errors for data qubits, as well as the coherent bit-flip error for the ancillary qubit. As one will see in later parts, these detection-induced coherent errors will result in undetected errors, which will accumulate into logical errors after a couple rounds of syndrome measurements. 
Similarly, for site operators, the extra term yields
\begin{equation}
\begin{split}
     &\quad\operatorname{H}_a\operatorname{CNOT}^{\prime}_{ad_4}\operatorname{CNOT}^{\prime}_{ad_3}\operatorname{CNOT}^{\prime}_{ad_2}\operatorname{CNOT}^{\prime}_{ad_1}\operatorname{H}_a|\Psi\rangle\\
     &\sim\left\{\mathbf{I}_a\left[(1-i\pi\kappa)\mathbf{I}^{\otimes 4}+\frac{i\pi\kappa}{4}(X_{d_1}+X_{d_2}+X_{d_3}+X_{d_4})\right]\right.\\
     &\quad-\left.X_a\left[-i\pi\kappa\mathbf{I}^{\otimes 4}+\frac{i\pi\kappa}{4}(X_{d_1}+X_{d_2}+X_{d_3}+X_{d_4})\right]\vphantom{\frac{i\pi\kappa}{4}}\right\}\\
     &\quad\times\operatorname{H}_a\operatorname{CNOT}_{ad_4}\operatorname{CNOT}_{ad_3}\operatorname{CNOT}_{ad_2}\operatorname{CNOT}_{ad_1}\operatorname{H}_a|\Psi\rangle.
\end{split}
\end{equation}
Thus, for site operator measurements, imprfect $\operatorname{CNOT}$ gates will introduce coherent $X$ errors for data qubits.


\section{Effect of modified stabilizer measurements and its alleviation}

Now we discuss the effect of those modified stabilizer measurements for quantum computation. Suppose the initial state is prepared perfectly in the logical code space, and then experiences error $E$. We further apply stabilizer measurements to detect errors. 
Here, we analyze effects of the detection-induced coherent error using the two-round strategy~\cite{versluis,brien}---first measuring plaquette (site) operators and followed by measuring site (plauqette) operators. Assume that the plaquette measurement is applied first, and direct calculations yield (see App.~\ref{sec:MS} for details) 
\begin{equation}
\begin{split}
    \mathcal{G}_Z^1(\operatorname{CNOT}^{\prime})E|\Psi\rangle&=\mathcal{D}^1(Z)\mathcal{G}_Z^1(\operatorname{CNOT})E|\Psi\rangle\\
    &=E^Z\left[\mathcal{D}^1(Z)E^X\mathcal{A}(E^X)\right]|\Psi\rangle,
\end{split}
\end{equation}
where $\mathcal{G}_{Z}^{1}\left(\operatorname{CNOT}^{\prime}\right)$ denotes measurements of {\em all} plaquette operators in the first round based on imperfect $\operatorname{CNOT}$ gates,  $E^{X}$ is the $X$ error acting on data qubits, operators acting on ancillary qubits and containing the information of $E^X$ is denoted by  $\mathcal{A}\left(E^{X}\right)$, and $\mathcal{D}^1(Z)$ is the total deviation from ideal plaquette measurements caused by imperfect $\operatorname{CNOT}$ gates in the first round. After the plaquette measurement, one should measure ancillary qubits so as to get the syndrome for $X$ errors. We use $\sum_{j}\mathcal{D}_{ij}^{1}\left(Z\right)$ to denote the term that survives after measurements of ancillary qubits, where $i$ stands for the $i$th independent ancilla configuration in $\mathcal{D}^{1}(Z)$, and the summation (over $j$) contains all coherent terms that share the same ancilla configuration. Thus, after the ancilla measurement, we have 
\begin{equation}
\begin{split}
    \mathcal{G}_Z^1(\operatorname{CNOT}^{\prime})E|\Psi\rangle\rightarrow \left[E^Z\sum_j\mathcal{D}_{ij}^1(Z)\right]E^X\mathcal{A}(E^X)|\Psi\rangle.  
\end{split}
\end{equation}
$E^Z\sum_j\mathcal{D}_{ij}^1(Z)$ contains $Z$ errors that need to be detected in the next round of site operator measurements, which leads to the following expression 
\begin{equation}
\begin{split}
    &\mathcal{G}_X^2(\operatorname{CNOT}^{\prime})\left[E^Z\sum_j\mathcal{D}_{ij}^1(Z)\right]E^X\mathcal{A}(E^X)|\Psi\rangle=\mathcal{D}^2(X)\\
    &\times\left[E^Z\mathcal{A}(E^Z)\right]\left[\sum_j\mathcal{D}_{ij}^1(Z)\mathcal{A}(\mathcal{D}_{ij}^1(Z))\right]\left[E^X\mathcal{A}(E^X)\right]|\Psi\rangle.
\end{split}
\end{equation}
Since each $\mathcal{D}_{ij}^{1}\left(Z\right)$ is an independent operator, each $\mathcal{A}\left(\mathcal{D}_{ij}^{1}\left(Z\right)\right)$ is also independent, which uniquely reflects the Z error configuration in $\mathcal{D}_{ij}^{1}\left(Z\right)$. After implementing this round of site operator measurements, we also need to measure ancillary qubits of site operators. Since $\mathcal{D}^{2}\left(X\right)$ also contains several ancilla configurations, $\mathcal{D}^{2}\left(X\right)\sum_{j}\mathcal{D}_{ij}^{1}\left(Z\right)\mathcal{A}\left(\mathcal{D}_{ij}^{1}\left(Z\right)\right)$ will result in some different Z errors sharing the same ancilla configuration, and this means the error correction can not exactly correct errors, and even introduce more errors. 
Therefore, the final state will be a coherent state with some correct terms and some incorrect terms. Those incorrect terms get worse due to the error correction procedure and will accumulate errors. Here, we suppose that all ancilla measurements are  error-free, and also ignore long error chains which is roughly longer than one half of the code size.
By doing so, error $E$ can be perfectly corrected, and we can concentrate on effects of the detection-induced coherent error just as there was no error $E$. In order to have a quick and clear interpretation of this effect, we repeatedly execute the quantum error correction cycles (including error detections and error corrections). Then, we use Mathematica based on the package {QUANTUM}~\cite{muoz} to simulate detection, decoding and error-correction procedures for the $d=3$ surface code according to the previous discussion (see App.~\ref{sec:MS} for details), we find undetected error terms accumulate errors and evolve to logical operators with a certain probability after several QEC cycles. For simplicity, we again keep terms up to the first order of imperfect rate $\kappa$. Similar analyses can be applied to other code sizes and a quantum circuit with logical operations. 



Calculating the probability of undetected error terms, that evolve to logical errors, is very hard for arbitrary code sizes, thus we can not exactly measure the performance of the surface code.
Fortunately, we can obtain the worst case performance. Note that the total deviation $\mathcal{D}(Z)$ contains several different configurations of operators acting on data qubits and ancillary qubits, such as the error-free configuration $\mathbf{I}^{\otimes (n+n_a^z)}$ with $n$ being the number of data qubits and $n_a^z$ being the number of ancillary qubits for plaquette operator measurements, and other configurations represent $Z$ errors on data qubits and $X$ errors on ancillary qubits induced by imperfect $\operatorname{CNOT}$ gates. For those configurations containing $\mathbf{I}^{\otimes n_a^z}$, there are $n+1$ possibilities, where one possibility corresponds to $\mathbf{I}^{\otimes(n+n_a^z)}$ and $n$ possibilities correspond to $\mathbf{I}^{\otimes(n-1+n_a^z)}\otimes Z_i$ with $i$ being the qubit acted by $Z$. The structure of $\mathcal{D}(X)$ is similar with $\mathcal{D}(Z)$. In order to simplify the analysis, we consider a worst-case scenario: only the case, that each round of ancilla measurements projects ancillary qubits into the state acted by $\mathbf{I}^{\otimes n_a}$, does not exist accumulated logical errors. For other cases (projecting to other ancilla configurations), we assume that the data qubits will finally experience accumulated logical errors (not all cases in reality). 

According to the structure of $\mathcal{D}$, we find the probability of projecting ancillary qubits into the state acted by $\mathbf{I}^{\otimes n_a}$ in $k$th round stabilizer measurements, under the condition that the $(k-1)$th round of ancilla measurements projects ancilla into the state acted by $\mathbf{I}^{\otimes n_a}$, is (see App.~\ref{sec:WCA} for details)
\begin{equation}
\begin{split}
    &\quad P_{k|k-1}(\mathbf{I}^{\otimes n_a})\\
    &=\mathcal{N}_k\left[1+\frac{k^2(2d^2-3d+1)^2}{4}\pi^2\kappa^2+\frac{4d^2-7d+2}{8}\pi^2\kappa^2\right],
\end{split}
\end{equation}
where $P_{1|0}\equiv P_1$, $\mathcal{N}_k$ is the normalized factor, and $k\ge1$ is the round count of stabilizer measurements (for the first plaquette round, $k=1$, and for the first site round, $k=2$, etc.). Combining with those cases that are not $\mathbf{I}^{\otimes n_a}$ configurations, we can derive the normalized factor (see App. \ref{sec:WCA} for the explicit expression).
After $2m$ rounds of syndromes measurements, the probability of the case, where each round of ancilla measurements projects ancilla into the state acted by $\mathbf{I}^{\otimes n_a}$, is
\begin{equation}
    P_{1\rightarrow 2m}(\mathbf{I}^{\otimes n_a})=P_{1}(\mathbf{I}^{\otimes n_a}) P_{2|1}(\mathbf{I}^{\otimes n_a})\cdots P_{2m|2m-1}(\mathbf{I}^{\otimes n_a}).
\end{equation}
For the worst case, the fidelity of the final state satisfies (see App. \ref{sec:WCA} for details)
\begin{equation}\label{eq:fidelity}
\begin{split}
    F&\ge P_{1\rightarrow 2m}(\mathbf{I}^{\otimes n_a})|\langle\Psi|\Psi_{f_1}\rangle|\\
    &= P_{1\rightarrow 2m}(\mathbf{I}^{\otimes n_a})|\alpha|,
\end{split}    
\end{equation}
where $|\Psi_{f_1}\rangle$ is the final state without logical errors, and $\alpha$ is the amplitude of the correct state in $|\Psi_{f_1}\rangle$ with 
\begin{equation}
   |\alpha|=\sqrt{\frac{1+m^2(2d^2-3d+1)^2\pi^2\kappa^2}{1+m^2(2d^2-3d+1)^2\pi^2\kappa^2+\frac{4d^2-7d+2}{8}\pi^2\kappa^2}}.
\end{equation}
In Eq.~\eqref{eq:fidelity}, we can focus on the minimum fidelity $F_{min}=P_{1\rightarrow 2m}(\mathbf{I}^{\otimes n_a})|\alpha|$, which characterizes the share of correct terms in the final state, to measure the performance of the surface code under detection-induced coherent errors. 

\begin{figure}
    \centering
    \includegraphics[height=5.5cm]{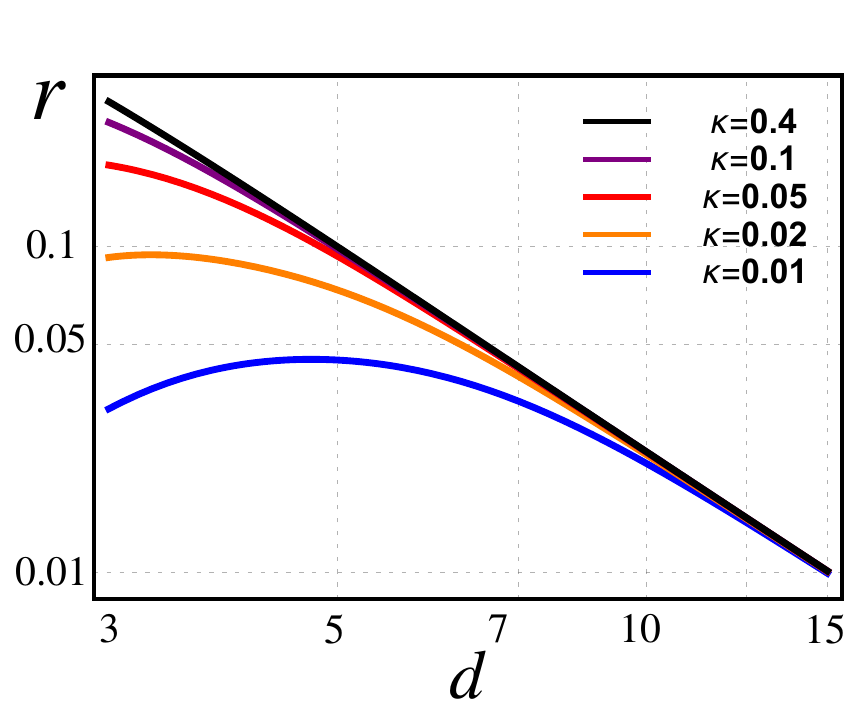}
    \caption{Log-log plot of the infidelity $r$ changing with the code size $d$. The infidelity is defined as $r=1-F_{min}$, where $F_{min}$ is the minimum fidelity. In this case, we take $m=3$.}
    \label{fig:infidelity}
\end{figure}

We plot the infidelity $r=1-F_{min}$ as a function of the code size $d$, at different imperfect rate $\kappa$s, shown in Fig.~\ref{fig:infidelity}. We can find that the infidelity in the worst case decreases with the increasing of the code size $d$ and decays in the form of power law functions in the large code size regime. Thus, the detection-induced coherent error can be alleviated by increasing $d$ (see App. \ref{sec:WCA} for discussions). We also find that with the increasing of $\kappa$ (still in the small $\kappa$ regime to ensure the small parameter expansion), the decay behavior saturates to $r\sim10^{0.47}d^{-2.1}$ (the top straight line in Fig.~\ref{fig:infidelity}). In Fig.~\ref{fig:infidelity}, we take $m=3$. For larger $m$s, $P_{1\rightarrow 2m}(\mathbf{I}^{\otimes n_a})$ will be smaller and $r$ will be larger. However, since for large $k$s, $P_{k|k-1}(\mathbf{I}^{\otimes n_a})$ will approach to $1$, and then there will be less differences between large-$m$ cases and small-$m$ cases (see App.~\ref{sec:WCA} for details). One may notice that for small code sizes, there is a bending behavior in the $r\sim d$ curve, indicating that in the error model under considerations, some larger-size codes  have worse performance than smaller-size codes. What's more, one may also notice large values of the infidelity of such errors. However, we emphasize that since we just focus on the worst case, negative results can not be used to judge QECCs. On the contrary, positiveness shown here again indicates the corresponding effectiveness of QECCs.


\section{Connection with approximate QECCs}\label{sec:CwAQECCs}
Before we discuss the connection between surface codes under the DICE and {\em approximate} QECCs, we provide a brief introduction to the Knill-Laflamme condition for the approximate QECCs which have attracted many interests~\cite{crepeau,beny,wangPRL18,faist,brandao}. For details, we refer to Ref. \cite{beny,brandao}.

Briefly speaking, approximate QECCs are QECCs which can not correct errors perfectly. Rigorously speaking, a QECC is called an approximate QECC, if it satisfies the Knill-Laflamme condition \cite{brandao}:
\begin{equation}\label{eq:KLCfAQECC}
    \langle\psi_i|O|\psi_j\rangle=C_{O}\delta_{ij}+\varepsilon_{ij}(O),
\end{equation}
where $\{|\psi_1\rangle,\cdots,|\psi_{2^k}\rangle\}$ are codewords that span the code space $\mathcal{C}$, and $O=E_a^{\dagger}E_b$ with $\{E_a\}$ being the operation elements of correctable local error channels $\mathcal{E}$. Note that the Kill-Laflamme condition for the approximate QECCs has an extra term $\varepsilon_{ij}(O)$ comparing with that of the exact QECCs. Intuitively, this tail implies that even local error operators can result in logical errors. If we let $i\neq j$, we have $\langle\psi_i|O|\psi_j\rangle$=$\varepsilon_{ij}(O)\neq 0$, and this tells us that the transition probability (transit through local operator $O$) from one logical state $|\psi_i\rangle$ to another logical state $|\psi_j\rangle$ is not zero. Note that in previous works, stabilizer measurements are assumed to be perfect, thus the encoding process using stabilizer operators is also perfect. Then, in those case, correctable errors can be perfectly corrected, and different logical states can not be converted through local operators. Therefore, those cases are within the category of {\em exact} QECCs. However, as we will see in the following, if stabilizer measurements are assumed to be imperfect due to the experimental limitation, situations will change, and approximate QECCs will come out naturally. 


We now return to our focus -- surface codes. The logical ``0'' state of the surface code can be written as \cite{kitaev,pachos}
\begin{equation}\label{0state}
    |\bar{0}\rangle=\prod_{s}\frac{1}{\sqrt{2}}(1+A(s))|0\rangle^{\otimes n},
\end{equation}
where $A(s)$ stands for site operator $X^{\otimes 4}$ and the logical ``1'' state can be derived from $|\bar{1}\rangle=X_L|\bar{0}\rangle$. For preparing $|\bar{0}\rangle$, we just need to implement one round of site operator measurements, then select an arbitrary open chain whose boundary consists of positions of all ancillary qubits with outcome $-1$ and apply $Z$ to each link of this chain \cite{dennis}. Previously, we assumed that the initial state can be prepared perfectly in the code space, but with the detection-induced coherent error, the initial state can not be perfect anymore. For example, practically, the logical ``0'' state and the logical ``1'' state will be dressed by $\mathcal{D}(X)$. Ancilla measurements will be more probable to project ancillary qubits into the state acted by $\mathbf{I}^{\otimes n_a}$ coming from $\mathcal{D}(X)$. Thus, the logical states in reality will be more likely to be $|\bar{0}^{\prime}\rangle=\mathcal{Y}(X)|\bar{0}\rangle$ and $|\bar{1}^{\prime}\rangle=\mathcal{Y}(X)|\bar{1}\rangle$, where for the $d=3$ surface code and keeping terms up to the first order of $\kappa$, 
\begin{equation}
\begin{split}
    \mathcal{Y}(X)&\propto\left(1-5i\pi\kappa\right)\mathbf{I}^{\otimes13}\\
    &\quad+\frac{i\pi\kappa}{4}\left(X_{1}+X_{2}+X_{3}+X_{11}+X_{12}+X_{13}\right)\\
    &\quad+\frac{i\pi\kappa}{2}\left(X_{4}+X_{5}+X_{6}+X_{7}+X_{8}+X_{9}+X_{10}\right),
\end{split}
\end{equation}
where we omit the normalization factor. For a general $d$, $\mathcal{Y}(X)$ has the same structure. Note that for the $d=3$ surface code, the correctable error $\mathcal{E}$ can be constructed from the operation elements $E_a=\{I,X_1\cdots,X_{13},Z_1\cdots,Z_{13},Y_1,\cdots,Y_{13}\}$. Therefore, taking $O=X_2$ as an example, we have $\langle\bar{1}^{\prime}|X_2|\bar{0}^{\prime}\rangle\propto\langle\bar{1}|\mathcal{Y}^{\dagger}(X)X_2\mathcal{Y}(X)|\bar{0}\rangle=(1/8)\pi^2\kappa^2\neq 0$. Note that the $d=3$ surface code $\left[\left[ 13,1,3\right]\right]$ is an  exact QECC, satisfying Kill-Laflamme condition $\langle\phi_i|O|\phi_j\rangle=C_O\delta_{ij}$ with $\left\{|\phi_i\rangle\right\}$ forming the code space of a QECC and $E$ being the correctable local errors \cite{knill}. However, above discussions show that due to the detection-induced coherent error, $\langle\phi_i^{\prime}|O|\phi_j^{\prime}\rangle\neq 0$ for $i\neq j$, which means the Knill-Laflamme condition becomes $\langle\phi_i^{\prime}|O|\phi_j^{\prime}\rangle=C_O\delta_{ij}+\varepsilon_{ij}(O)$. One finds that this is the Knill-Laflamme condition for  approximate QECCs, Eq. \eqref{eq:KLCfAQECC}.  For the $d=3$ suface code case, the exact structure of $\varepsilon_{ij}(O)$ can be found in App. \ref{sec:DSV}.  Note that for a general $d$, the structure of $\varepsilon_{ij}(O)$ has the same form with modifications of values, and the derivation is also similar with that of the $d=3$ case.

It is easy to check that considering larger code sizes, some local errors still satisfies the exact Knill-Laflamme condition, if we just keep terms up to the first order of $\kappa$. 
However, the detection-induced coherent error can also introduce higher order terms, thus in general surface codes become approximate QECCs. In order to justify the whole QEC procedure, we need to include QEC cycles (as shown in the last section) along with the initial encoding step. Since this encoding step just needs one round of site operator measurements, our former discussions can be applied immediately and imply that the residual errors of approximate QECCs
can be alleviated. Previous studies show that allowing negligible errors existing in the recovery procedure can lead to better QECCs \cite{benaroya,leung,crepeau}, thus together with our result, one can anticipate that there may be good approximate QECCs suitable for fault-tolerant quantum computation.


\section{Summary and discussion}

Based on the surface code, we show that the detection-induced coherent error will result in accumulated logical errors after running a quantum circuit, and this impact can be alleviated by increasing the code size.  Effects of the detection-induced coherent error may be more serious for a nontrivial quantum circuit beyond QEC-only circuits, and we leave it to further investigations. We also show that the actual logical states are imperfect under the detection-induced coherent error, and then the exact surface code becomes an approximate QECC. Therefore, our results imply that some approximate QECCs may also be possible for fault-tolerant quantum computation.

\begin{acknowledgments} 
 Authors thank Xuegang Li, Yuanchen Zhao, Li Rao, Rui Chao, and Xiaosi Xu for helpful discussions. The work is supported by NSF-China (GrantNo.11974198) and the startup grant from State Key Laboratory of Low-Dimensional Quantum Physics of Tsinghua University. 
\end{acknowledgments}

\begin{appendix}
\begin{widetext}
\section{Connection between imperfect rate and minimum gate fidelity}\label{sec:IR&MGF}
As mentioned in the main text, the imperfect CNOT gate can be modeled by $\operatorname{CNOT}=\operatorname{H}_t\Phi^{\kappa}_{ct}\Phi_{ct}\operatorname{H}_t$. Since the imperfect rate $\kappa$ is a theoretical quantity and can not be directly measured experimentally, we need to connect it with an experimentally measurable quantity, and then we can know the performance of a CNOT gate with a specific imperfect rate. Here we show the connection between the imperfect rate and the minimum gate fidelity, defined as~\cite{nielsen}
\begin{equation}
    F_G(U,\mathcal{E})\equiv \min_{|\psi\rangle}F(U|\psi\rangle,\mathcal{E}(|\psi\rangle\langle\psi|)),
\end{equation}
where $U$ stands for the operation of an ideal quantum gate, $\varepsilon$ stands for the quantum channel representing an imperfect quantum gate, $F$ is the fidelity between two quantum states, and the minimization is over all possible two-qubit states. Here we rewrite the ideal CNOT as $U$ and the imperfect CNOT as $U^{\prime}$, then the minimum gate fidelity is 
\begin{equation}\label{eq:MGF}
\begin{split}
    F_G(U,U^{\prime})&=\min_{|\psi\rangle}F(U|\psi\rangle,U^{\prime}|\psi\rangle)\\
    &=\min_{|\psi\rangle}\operatorname{Tr}\sqrt{\left(U|\psi\rangle\langle\psi|U^{\dagger}\right)^{1/2}\left(U^{\prime}|\psi\rangle\langle\psi|U^{\prime\dagger}\right)\left(U|\psi\rangle\langle\psi|U^{\dagger}\right)^{1/2}}\\
    &=\min_{|\psi\rangle}|\langle\psi|UU^{\prime}|\psi\rangle|\\
    &=\min_{|\psi\rangle}|a+be^{-i\pi\kappa}|\\
        &=\min_{|\psi\rangle}\left(a\sqrt{1+2\frac{b}{a}\cos\left(\pi\kappa\right)+\left(\frac{b}{a}\right)^{2}}\right)\\
        &=\frac{1}{2}\sqrt{2+2\cos\left(\pi\kappa\right)},
\end{split}
\end{equation}
where 
\begin{equation}
    \begin{split}
        a&=\frac{3}{4}+\frac{1}{4}\left[\left(|\alpha|^{2}+|\beta|^{2}-|\gamma|^{2}-|\kappa|^{2}\rangle\right)+\left(2\gamma^{*}\kappa+2\gamma\kappa^{*}\right)\right],\\b&=\frac{1}{4}-\frac{1}{4}\left[\left(|\alpha|^{2}+|\beta|^{2}-|\gamma|^{2}-|\kappa|^{2}\rangle\right)+\left(2\gamma^{*}\kappa+2\gamma\kappa^{*}\right)\right],
    \end{split}
\end{equation}
we have let $|\psi\rangle=\alpha|0\rangle_{c}|0\rangle_{t}+\beta|0\rangle_{c}|1\rangle_{t}+\gamma|1\rangle_{c}|0\rangle_{t}+\kappa|1\rangle_{c}|1\rangle_{t}$ with $\alpha,\beta,\gamma$ and $\gamma$ being complex numbers, and the minimum is obtained when $\left(|\alpha|^{2}+|\beta|^{2}-|\gamma|^{2}-|\kappa|^{2}\rangle\right)+\left(2\gamma^{*}\kappa+2\gamma\kappa^{*}\right)=-1$. For this minimum condition, $\alpha=\beta=\gamma=0,\kappa=1$ and $\alpha=\beta=\kappa=0,\gamma=1$ are possible configurations, and the corresponding states are $|\psi\rangle=|1\rangle_c|0\rangle_t$ and $|\psi\rangle=|1\rangle_c|1\rangle_t$. Since they are both direct product states, they can be easily prepared experimentally(if we measure the minimum gate fidelity of imperfect controlled-Z gate, the state statisfying the minimum condition is a coherent state, which will be difficult for the experimental preparation), and then one can measure the minimum gate fidelity instead of the average gate fidelity through randomized benchmarking. According to Eq.~\eqref{eq:MGF}, one can know that if $\kappa=0.01,0.02,0.05,0.1$ and $0.4$(values shown in Fig.~2 in the main text), then the minimum gate fidelity will be $0.99988,0.99951,0.9969,0.9877$ and $0.809$, respectively.

\section{Notes on the simulation of $d=3$ surface code }\label{sec:MS}
As mentioned in the main text, we use the two-round strategy to implement the simulation. We first do the plaquette measurement, whose operation can be expressed as 
\begin{equation}
\begin{split}
    \mathcal{G}_Z^1(\operatorname{CNOT}^{\prime})E|\Psi\rangle&=\mathcal{D}^1(Z)\mathcal{G}_Z^1(\operatorname{CNOT})E|\Psi\rangle\\
    &=\mathcal{D}^1(Z)\mathcal{G}_Z^1(\operatorname{CNOT})E^ZE^X|\Psi\rangle\\
    &=E^Z\mathcal{D}^1(Z)\left[\mathcal{G}_Z^1(\operatorname{CNOT})E^X\mathcal{G}_Z^1(\operatorname{CNOT})\right]\mathcal{G}_Z^1(\operatorname{CNOT})|\Psi\rangle\\
    &=E^Z\left[\mathcal{D}^1(Z)E^X\mathcal{A}(E^X)\right]|\Psi\rangle,
\end{split}
\end{equation}
in the second equality, we split $E$ into $Z$ errors and $X$ errors, and in the forth equality, we have used the fact that $|\Psi\rangle$ is an perfect state in the code space and is invariant under the action of ideal stabilizers. $\mathcal{A}(E^X)$ results from the error propagation nature of $\operatorname{CNOT}$ gates, which will propagate $X$ errors from data qubits to ancillary qubits. Attributed to this feature, one can figure out errors in data qubits from syndromes shown in ancillary qubits. $\mathcal{D}^1(Z)$ is the total deviation from ideal plaquette measurements caused by imperfect $\operatorname{CNOT}$ gates in the first round, and can be represented as $\mathcal{D}^1(Z)=\sum_{i}\sum_j\mathcal{D}^1_{ij}(Z)$, where $i$ stands for the $i$th independent ancilla configuration, and $j$ stands for $j$th term that shares the same ancilla configuration. The expression of $\mathcal{D}^{1}(Z)$, which is the product of deviations of all plaquette stabilizers(Eq.~(5) in the main text), reads as 
\begin{equation}
\begin{split}\label{eq:d1z}
    \mathcal{D}^{1}(Z)=&\left[(1-5i\pi\kappa)\mathbf{I}^{\otimes 13}  +\frac{1}{4}i\pi\kappa\left(Z_{1}+Z_{3}+Z_{6}+Z_{8}+Z_{11}+Z_{13}\right)
    +\frac{1}{2}i\pi\kappa\left(Z_{2}+Z_{4}+Z_{5}+Z_{7}+Z_{9}+Z_{10}+Z_{12}\right)\right]\mathbf{I}^{\otimes 6}_a\\
    &\quad+\frac{1}{4}i\pi\kappa(3\mathbf{I}^{\otimes 13}+Z_1+Z_2+Z_4)X_{a1}+\frac{1}{4}i\pi\kappa(3\mathbf{I}^{\otimes 13}+Z_2+Z_3+Z_5)X_{a2}\\
    &\quad +\frac{1}{4}i\pi\kappa(3\mathbf{I}^{\otimes 13}+Z_9+Z_{11}+Z_{12})X_{a11}+\frac{1}{4}i\pi\kappa(3\mathbf{I}^{\otimes 13}+Z_{10}+Z_{12}+Z_{13})X_{a12}\\
    &\quad+\frac{1}{4}i\pi\kappa(4\mathbf{I}^{\otimes 13}+Z_4+Z_6+Z_7+Z_9)X_{a6}+\frac{1}{4}i\pi\kappa(4\mathbf{I}^{\otimes 13}+Z_5+Z_7+Z_8+Z_{10})X_{a7},
\end{split}
\end{equation}
where $a1,a2,a6,a7,a11$ and $a12$ stand for the ancillary qubits (as shown in Fig. \ref{fig:code_illustration} (b) of the main text) used for implementing plaquette operator measurements, $i(i=1,\cdots,13)$ stands for data qubits, and we have kept terms up to the first order of $\kappa$, 
and used the fact that for the three-operator stabilizers, such as $Z_{d_1}Z_{d_2}Z_{d_3}$, the deviation(Eq.(5) in the main text) reads as 
\begin{equation}
\begin{split}
    \operatorname{H}_a\Phi_{d_3a}^{\kappa}\Phi_{d_2a}^{\kappa}\Phi_{d_1a}^{\kappa}\operatorname{H}_a\sim\left[(1-\frac{3i\pi\kappa}{4})\mathbf{I}^{\otimes 4}+\frac{i\pi\kappa}{4}(Z_{d_1}+Z_{d_2}+Z_{d_3})\right]\mathbf{I}_a
    +\left[\frac{3i\pi\kappa}{4}\mathbf{I}^{\otimes 4}-\frac{i\pi\kappa}{4}(Z_{d_1}+Z_{d_2}+Z_{d_3})\right]X_a.
\end{split}
\end{equation}
Similar for site operators.

For a particular ancilla configuration, say $X_{a6}$, $\sum_j\mathcal{D}^1_{ij}(Z)$ reads as 
\begin{equation}
\begin{split}
   \sum_j\mathcal{D}^1_{ij}(Z)=\frac{1}{4}i\pi\kappa(4\mathbf{I}^{\otimes 13}+Z_4+Z_6+Z_7+Z_9)X_{a6},
\end{split}
\end{equation}
where $\mathbf{I}^{\otimes 13},Z_4,Z_6,Z_7$ and $Z_9$ are five different error configurations sharing the same ancilla configuration, which resulted from the imperfect implementation of $Z_4Z_6Z_7Z_9$ stabilizer.  After measuring the ancillary qubits, only one ancilla configuration survives. Thus, we have 
\begin{equation}
\begin{split}
    \mathcal{G}_Z^1(\operatorname{CNOT}^{\prime})E|\Psi\rangle\rightarrow \left[E^Z\sum_j\mathcal{D}_{ij}^1(Z)\right]E^X\mathcal{A}(E^X)|\Psi\rangle.  
\end{split}
\end{equation}
For the next site operator measurements, by similar analysis, we have 
\begin{equation}
\begin{split}
    &\mathcal{G}_X^2(\operatorname{CNOT}^{\prime})\left[E^Z\sum_j\mathcal{D}_{ij}^1(Z)\right]E^X\mathcal{A}(E^X)|\Psi\rangle=\mathcal{D}^2(X)\left[E^Z\mathcal{A}(E^Z)\right]\left[\sum_j\mathcal{D}_{ij}^1(Z)\mathcal{A}(\mathcal{D}_{ij}^1(Z))\right]\left[E^X\mathcal{A}(E^X)\right]|\Psi\rangle.
\end{split}
\end{equation}
After measuring site operators, we also need to measure the ancillary qubits to get the syndrome. Since there are also several ancilla configurations in $\mathcal{D}^2(X)$, $\mathcal{D}^{2}\left(X\right)\sum_{j}\mathcal{D}_{ij}^{1}\left(Z\right)\mathcal{A}\left(\mathcal{D}_{ij}^{1}\left(Z\right)\right)$ will result in some different Z error configurations sharing the same ancilla configuration, such as $1/4i\pi\kappa(3\mathbf{I}^{\otimes 13}+X_3+Z_3+X_5+X_8)X_{a5}$, where $a5$ is the ancilla for site operators used to detect $Z$ errors, and $X_3,X_5$ and $X_8$ errors are introduced from the imperfect implementation of $X_3X_5X_8$ stabilizer. If we correct $Z$ errors according to the syndrome, then we will apply $Z_3$ to the state of data qubits, and then what we will get is not a corrected state(without $Z$ errors), but a state acted by $(3Z_3+Y_3+\mathbf{I}^{\otimes 13}+Z_3X_5+Z_3X_8)$, which is a state with more $Z$ errors. Above analyses are based on keeping terms up to the first order of $\kappa$; and for  higher order terms, we have the same picture. Similar analysis show that with the proceeding of detection-correction processes, those incorrect terms will get worse and accumulate more errors, and will evolve into logical errors.

\section{Notes on the worst-case analysis}\label{sec:WCA}

Since there are too many possibilities, it is unlikely to calculate the probability of undetected error terms evolving to logical errors. Therefore, we can not exactly measure the performance of the surface code in such errors. Thus, we turn to the discussion of the worst case. We consider the worst case: Only for the case that each round of ancilla measurements projects ancillary qubits into the state acted by $\mathbf{I}^{\otimes n_a}$, there will be a chance with no accumulated logical errors. For other cases (projecting to other ancilla configurations), we assume that final data qubits will always experience accumulated logical errors (not all cases in reality). Thus, for measuring the performance of the surface code under the detection-induced coherent error, we just need the probability of projecting ancillary qubits into the state acted by $\mathbf{I}^{\otimes  n_a}$ in each round of stabilizer measurements---$P_{k|k-1}(\mathbf{I}^{\otimes n_a})$. We now explain how to derive $P_{k|k-1}(\mathbf{I}^{\otimes n_a})$.

\begin{figure}
    \centering
    \includegraphics[height=7cm]{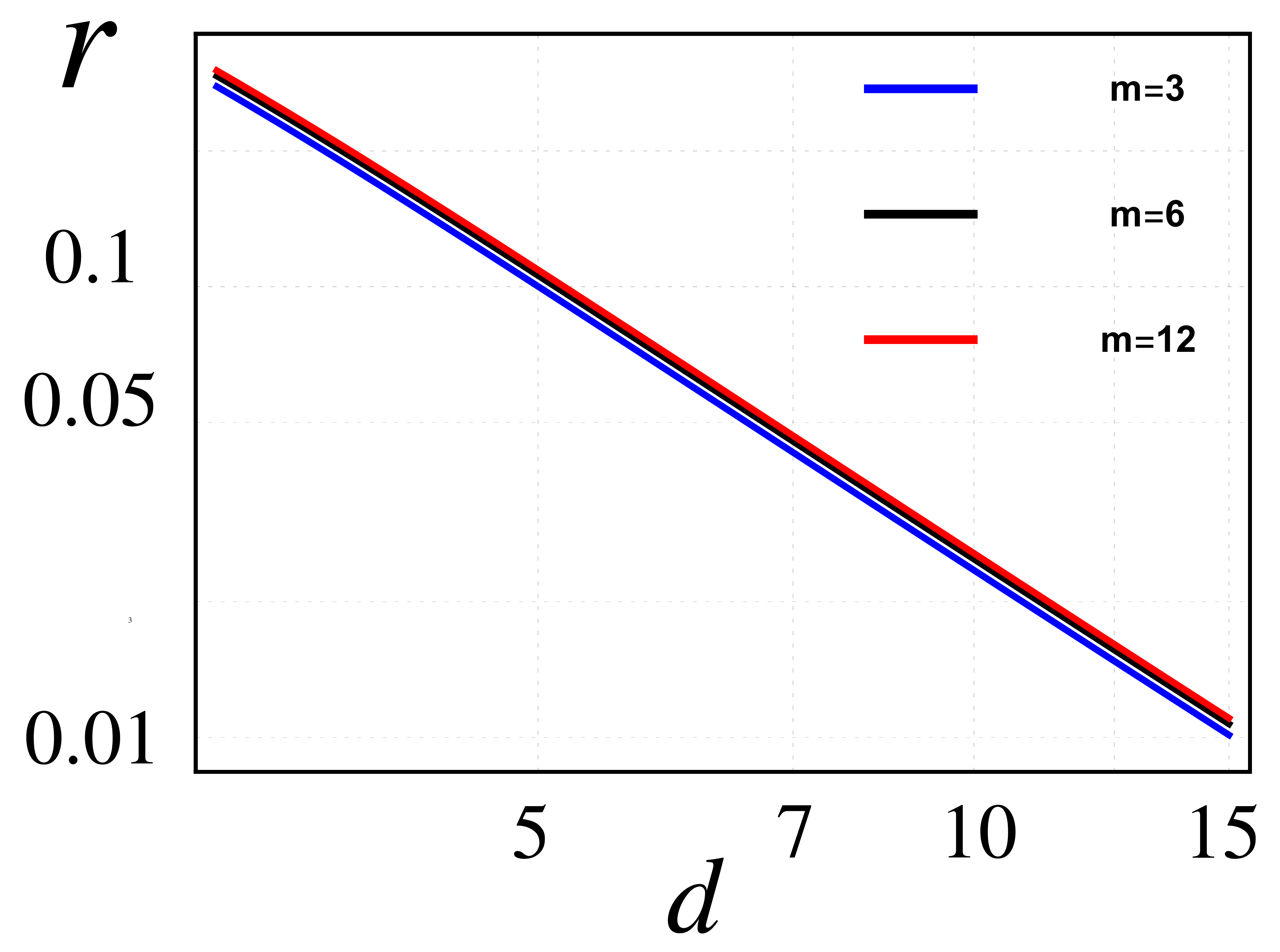}
    \caption{Log-log plot of the infidelity $r$ changing with the code size $d$ for different $m$'s. $\kappa$ is chosen to be $0.4$ here. One can find that the infidelity (for the worst case) is insensitive to $m$.}
    \label{fig:infidelity_for_ms}
\end{figure}

From the configuration of the surface code, we know that for one specific round of stabilizer measurements, there are 
\begin{enumerate}
    \item $4$ data qubits, each of which only belongs to one three-operator stabilizer;
    \item $2\times(d-2)$ data qubits, each of which only belongs to one four-operator stabilizers;
    \item $2\times(d-2)$ data qubits, each of which is shared by two three-operator stabilizers;
    \item $2\times(d-1)$ data qubits, each of which is shared by one three-operator stabilizer and one four-operator stabilizer;
    \item $(d-2)^2+(d-3)(d-1)$ data qubits, each of which is shared by two four-operator stabilizers.
\end{enumerate}

Since $\mathcal{D}^k$ is the product of the deviation of each stabilizer measurement, one can get the amplitude of each independent term in $\mathcal{D}^{k}$(see Eq. \eqref{eq:d1z} for $\mathcal{D}^{1}(Z)$ of the $d=3$ case). For example, the amplitude of the configuration $\mathbf{I}^{\otimes (n+n_a)}$ is $1-k(2d^2-3d+1)i\pi\kappa/2$ with $k$ being the round count of stabilizer measurements, and the amplitude of the configuration $\mathbf{I}^{\otimes (n-1+n_a^z)}\otimes Z_1$ is $i\pi\kappa/4$(since qubit-$1$ only belongs to one three-operator stabilizer, $Z$ errors occurring on it can only come from the deviation of this stabilizer, and then only one $i\pi\kappa/4$ factor will contribute to the amplitude.). Amplitudes of other configurations can also be derived from the same strategy. Thus, from those amplitudes, one can get the probability of projecting ancillary qubits into the state acted by $\mathbf{I}^{\otimes n_a}$ in $k$th round stabilizer measurements under the condition that the $(k-1)$th round of ancilla measurements projects ancilla into the state acted by $\mathbf{I}^{\otimes n_a}$:
\begin{equation}
\begin{split}
    P_{k|k-1}(\mathbf{I}^{\otimes n_a})=\mathcal{N}_k\left[1+\frac{k^2(2d^2-3d+1)^2}{4}\pi^2\kappa^2+\frac{4d^2-7d+2}{8}\pi^2\kappa^2\right],
\end{split}
\end{equation}
where 
\begin{equation}
    \mathcal{N}_k=\frac{1}{1+\frac{k^2(2d^2-3d+1)^2}{4}\pi^2\kappa^2+(1+\theta(k-2))\frac{4d^2-7d+2}{8}\pi^2\kappa^2+\frac{5d^2-9d+4}{4}\pi^2\kappa^2}
\end{equation}
with $\theta(x)$ being the unit step function, and  $\theta(x\ge 0)=1$ while $\theta(x<0)=0$. The $\theta$ function comes from the fact that the first round stabilizer measurements is different from the subsequent ones, as the first round stabilizer measurements measures a perfect state. The forth term in the denominator of $\mathcal{N}_k$ comes from those configurations that ancillary qubits are not acted by $\mathbf{I}^{\otimes n_a}$. One should note that for calculating the amplitude, we keep terms up to the first order of $\kappa$, and then we should keep terms up to $\kappa^2$ for the probability.

For the fidelity of the final state after $2m$ rounds of stabilizer measurements in the worst case, we know that there will be three possibilities of logical errors for the surface code, that is $X_L,Z_L$ and $X_LZ_L$. For a general logical state $|\psi\rangle=a|\bar{0}\rangle+b|\bar{1}\rangle$, if it experiences logical errors, then the fidelity will be $F=|\langle \Psi|E_L|\Psi\rangle|\ge 0$, where $E_L$ denotes the logical error. For example, if $|\Psi\rangle=|\bar{0}\rangle$ and $E_L=X_L$, then $F=0$. Since there will exist three possibilities of logical errors with different probabilities in the worst case, the fidelity of the final state satisfies $F\ge P_{1\rightarrow 2m}(\mathbf{I}^{\otimes n_a})|\langle\Psi|\Psi_{f_1}\rangle|$, where we have used the fact that the minimum of the fidelity between the original state and  the final state with logical errors is $0$, and $|\Psi_{f_1}\rangle$ is the final state without logical errors and its probability is $P_{1\rightarrow 2m}(\mathbf{I}^{\otimes n_a})$ in the worst case. Note that even though each round of ancilla measurements projects ancillary qubits into the state acted by $\mathbf{I}^{\otimes n_a}$, data qubits can still suffer from errors (such as there will be $\mathbf{I}^{\otimes (n-1+n_a^z)}\otimes Z_i$ in $\mathcal{D}(Z)$). Thus, $|\Psi_{f_1}\rangle$ can be expressed as $|\Psi_{f_1}\rangle=\alpha|\Psi\rangle+|\Psi^{\prime}\rangle$, where $\alpha$ is the amplitude of the correct state in $|\Psi_{f_1}\rangle$ and $|\Psi^{\prime}\rangle$ is a state with non-logical errors. Then, $|\langle\Psi|\Psi_{f_1}\rangle|=|\alpha+\langle\Psi|\Psi^{\prime}\rangle|=|\alpha|$. Therefore, $F\ge P_{1\rightarrow 2m}(\mathbf{I}^{\otimes n_a})|\langle\Psi|\Psi_{f_1}\rangle|=P_{1\rightarrow 2m}(\mathbf{I}^{\otimes n_a})|\alpha|$, and 
\begin{equation}
   |\alpha|=\sqrt{\frac{1+m^2(2d^2-3d+1)^2\pi^2\kappa^2}{1+m^2(2d^2-3d+1)^2\pi^2\kappa^2+\frac{4d^2-7d+2}{8}\pi^2\kappa^2}}.
\end{equation}

As mentioned in the main text, the infidelity $r=1-F_{min}$ decays with the increasing of code sizes. If we do not focus on the worst case, then the final state will be a coherent state containing correct terms, non-logical error terms and logical error terms. Since the actual fidelity of the final state $F$ is larger than the minimum fidelity $F_{min}$, which is contributed by the correct terms in the final state, the extra fidelity $F-F_{min}$ must come from the logical error terms(those non-logical error terms are orthogonal with the perfect state). If we focus on the worst case, then the final state will just be a coherent state containing correct terms and logical error terms(since $|\alpha|\rightarrow 1$, it is almost accurate to say this.). Then, suppose $F=1$(taking this maximum will facilitate the analysis, and the corresponding case can be that the initial state is $|\bar{0}\rangle$ with the logical error being $Z_L$), and $1-F_{min}$ is the contribution from logical error terms. In the main text, we show that $F_{min}$ will increase by increasing code sizes($r$ decays), thus the share of the correct terms in the final state will get higher with the increasing of code sizes, and then the computation accuracy will be higher. Thus, we can say that the effect of the detection-induced coherent error can be alleviated by QECCs. 

In Fig. \ref{fig:infidelity} in the main text, we choose $m=3$. For larger $m$, $P_{1\rightarrow 2m}(\mathbf{I}^{\otimes n_a})$ will be smaller and $r$ will be larger. However, since for large $k$, $P_{k|k-1}(\mathbf{I}^{\otimes n_a})$ will approach to $1$, and then there will be less differences between large-$m$ cases and small-$m$ cases (see Fig.~\ref{fig:infidelity_for_ms}). One should note that this does not mean that the repeating round of stabilizer measurements does not have any effect. In reality(do not focus on the worst case), it is more possible to appear logical error terms in the final state with larger $m$. 
Therefore, at a fixed code size, even though the final state may have a high fidelity in the larger $m$ case, we can not perceive that the state is in good quality, as the logical error terms can also contribute to the fidelity. Therefore, the final computing results may not be reliable, although used states are in a high fidelity.

\section{Derivation of the exact structure of $\varepsilon_{ij}(O)$}\label{sec:DSV}
In Sec. \ref{sec:CwAQECCs} of the main text, based on an explicit example, we show that the surface code under the DICE becomes an approximate QECC, which satisfies the modified Knill-Laflamme condition
\begin{equation}
\langle\phi_{i}^{\prime}|O|\phi_{j}^{\prime}\rangle=C_{O}\delta_{ij}+\varepsilon_{ij}\left(O\right).    
\end{equation}
Since the surface code is a degenerate code, $C_{O}=1$ for $O=I$, and $C_{O}=0$ for $O\neq I$. We now show the exact structure of $\varepsilon_{ij}\left(O\right)$.

Note that for the d=3 surface code, the correctable error set is $\left\{ I,X_{1},\cdots,X_{13},Z_{1},\cdots,Z_{13},Y_{1},\cdots,Y_{13}\right\}$ . Since $O=E_{a}^{\dagger}E_{b}$, 
\begin{equation}
\begin{split}
    O&\in\left\{ I,X_{1},\cdots X_{13},X_{a}X_{b}\left(a,b=1,\cdots,13|a\neq b\right),Z_{1},\cdots,Z_{13},Z_{a}Z_{b}\left(a,b=1,\cdots,13|a\neq b\right),\right.\\
    &\quad\;\left.X_{a}Z_{b}\left(a,b=1,\cdots,13\right),X_{a}Y_{b}\left(a,b=1,\cdots,13\right),Y_{a}Z_{b}\left(a,b=1,\cdots,13\right)\right\} .
\end{split}
\end{equation} 
Those $O$s can be classified into four classes:

\begin{enumerate}
    \item $O=I$;
    
    \item The second one is constructed from $O$s containing $Z$ operators;
    
    \item The third one is constructed from $O$s containing only one $X$ operator;
    
    \item The last one is constructed from $O$s containing two $X$ operators.
\end{enumerate}

To avoid tedious algebraic calculations of $\langle\phi_{i}^{\prime}|O|\phi_{j}^{\prime}\rangle$, where $|\phi_{i}^{\prime}\rangle=|\bar{0}^{\prime}\rangle,|\bar{1}^{\prime}\rangle$, we provide some intuitive analyses. Note that $\langle\phi_{i}^{\prime}|O|\phi_{j}^{\prime}\rangle\propto\langle\phi_{i}|\mathcal{Y}^{\dagger}\left(X\right)O\mathcal{Y}\left(X\right)|\phi_{j}\rangle$, where $|\phi_{i}\rangle=|\bar{0}\rangle,|\bar{1}\rangle$. Thus, for $i=j$, in order that $\langle\phi_{i}|\mathcal{Y}^{\dagger}\left(X\right)O\mathcal{Y}\left(X\right)|\phi_{i}\rangle$ is not zero, $\mathcal{Y}^{\dagger}\left(X\right)O\mathcal{Y}\left(X\right)$ should be proportional to the identity or stabilizers; for $i\neq j$, in order that $\langle\phi_{i}|\mathcal{Y}^{\dagger}\left(X\right)O\mathcal{Y}\left(X\right)|\phi_{j}\rangle$ is not zero, $\mathcal{Y}^{\dagger}\left(X\right)O\mathcal{Y}\left(X\right)$ should be logical operators. Based on those intuitive analyses, $\langle\phi_{i}^{\prime}|O|\phi_{j}^{\prime}\rangle$ can be calculated very quickly, and we show the results in the following.

For $O=I$, we have
\begin{equation}
\langle\bar{0}^{\prime}|I|\bar{0}^{\prime}\rangle=\langle\bar{1}^{\prime}|I|\bar{1}^{\prime}\rangle\propto\langle\bar{0}|\mathcal{Y}^{\dagger}\left(X\right)I\mathcal{Y}\left(X\right)|\bar{0}\rangle=1,
\end{equation}
and 
\begin{equation}
    \langle\bar{0}^{\prime}|I|\bar{1}^{\prime}\rangle=\langle\bar{1}^{\prime}|I|\bar{0}^{\prime}\rangle\propto\langle\bar{0}|\mathcal{Y}^{\dagger}\left(X\right)I\mathcal{Y}\left(X\right)|\bar{1}\rangle=0.
\end{equation}
Therefore, for $O=I$, we have $\varepsilon_{ij}\left(O\right)=0$. 

We now consider those Os which contains Z operator. Simple algebraic calculations lead to 
\begin{equation}
\langle\bar{0}^{\prime}|Z_{a}Z_{b}|\bar{0}^{\prime}\rangle\propto\langle\bar{0}|\mathcal{Y}^{\dagger}\left(X\right)Z_{a}Z_{b}\mathcal{Y}\left(X\right)|\bar{0}\rangle=0,
\end{equation}
\begin{equation}
\langle\bar{1}^{\prime}|Z_{a}Z_{b}|\bar{1}^{\prime}\rangle\propto\langle\bar{0}|\mathcal{Y}^{\dagger}\left(X\right)X_{L}Z_{a}Z_{b}X_{L}\mathcal{Y}\left(X\right)|\bar{0}\rangle\propto\langle\bar{0}|\mathcal{Y}^{\dagger}\left(X\right)Z_{a}Z_{b}\mathcal{Y}\left(X\right)|\bar{0}\rangle=0,    
\end{equation}
\begin{equation}
    \begin{split}
        \langle\bar{0}^{\prime}|Z_{a}Z_{b}|\bar{1}^{\prime}\rangle&=\langle\bar{1}^{\prime}|Z_{a}Z_{b}|\bar{0}^{\prime}\rangle\\&\propto\langle\bar{0}|\mathcal{Y}^{\dagger}\left(X\right)Z_{a}Z_{b}X_{L}\mathcal{Y}\left(X\right)|\bar{0}\rangle\\&=\langle\bar{0}|\mathcal{Y}^{\dagger}\left(X\right)Z_{a}Z_{b}\mathcal{Y}\left(X\right)X_{L}|\bar{0}\rangle\\&=\langle\bar{0}|\mathcal{Y}^{\dagger}\left(X\right)Z_{a}Z_{b}\mathcal{Y}\left(X\right)|\bar{1}\rangle\\&=0,
    \end{split}
\end{equation}
\begin{equation}
    \langle\bar{0}^{\prime}|X_{a}Z_{b}|\bar{0}^{\prime}\rangle\propto\langle\bar{0}|\mathcal{Y}^{\dagger}\left(X\right)X_{a}Z_{b}\mathcal{Y}\left(X\right)|\bar{0}\rangle=0,
\end{equation}
\begin{equation}
    \langle\bar{1}^{\prime}|X_{a}Z_{b}|\bar{1}^{\prime}\rangle\propto\langle\bar{0}|\mathcal{Y}^{\dagger}\left(X\right)X_{L}X_{a}Z_{b}X_{L}\mathcal{Y}\left(X\right)|\bar{0}\rangle\propto\langle\bar{0}|\mathcal{Y}^{\dagger}\left(X\right)X_{a}Z_{b}\mathcal{Y}\left(X\right)|\bar{0}\rangle=0,
\end{equation}
and 
\begin{equation}
    \begin{split}
        \langle\bar{0}^{\prime}|X_{a}Z_{b}|\bar{1}^{\prime}\rangle&=\langle\bar{1}^{\prime}|X_{a}Z_{b}|\bar{0}^{\prime}\rangle\\&\propto\langle\bar{0}|\mathcal{Y}^{\dagger}\left(X\right)X_{a}Z_{b}X_{L}\mathcal{Y}\left(X\right)|\bar{0}\rangle\\&=\langle\bar{0}|\mathcal{Y}^{\dagger}\left(X\right)X_{a}Z_{b}\mathcal{Y}\left(X\right)|\bar{1}\rangle\\&=0.
    \end{split}
\end{equation}

Similarly, we have $\varepsilon_{ij}\left(Z_{a}\right)=0$. Therefore, we find that for those $O$s containing $Z$ operators, $\varepsilon_{ij}\left(O\right)$ always equals to $0$.

According to similar calculations, for the case that $O$ only contains one $X$ operator, we have 
\begin{itemize}
            \item $i\neq j$:\\
            \begin{equation}
            \begin{split}
            \varepsilon_{ij}(X_a) \propto\begin{cases}
            \frac{1}{8}\pi^{2}\kappa^{2} &  a=1,2,3,11,12,13\\
            0 &  a=4,5,9,10\\
            \frac{1}{2}\pi^{2}\kappa^{2} &  a=6,7,8
            \end{cases};
            \end{split}
            \end{equation}
            
            \item $i=j$:
            \begin{equation}
            \begin{split}
            \varepsilon_{ij}\left(X_{a}\right)\propto\begin{cases}
            -2\pi^{2}\kappa^{2} & a=1,3,11,13\\
            -\frac{19}{4}\pi^{2}\kappa^{2} & a=4,5,9,10\\
            -\frac{5}{2}\pi^{2}\kappa^{2} & a=2,12\\
            -\frac{9}{2}\pi^{2}\kappa^{2} & a=6,8\\
            -5\pi^{2}\kappa^{2} & a=7
            \end{cases}.
            \end{split}
            \end{equation}
        \end{itemize}
And for the case that $O$ contains two $X$ operators, we have 
\begin{itemize}
    \item $i\neq j$:
    \begin{equation}
        \varepsilon_{ij}\left(O=X_{a}X_{b}\right)\\
            \;
           \propto\begin{cases}
            -\frac{5}{2}\pi^{2}\kappa^{2} & a,b\in\left\{ 1,2,3,11,12,13\right\} \\
            & \cup\left(a,b\operatorname{in\;the\;same\;row}\right)\\
            -5\pi^{2}\kappa^{2} & a,b\in\left\{ 6,7,8\right\} \\
            0 & a,b\in\left\{ 1,2,3,6,7,8,11,12,13\right\} \\
            & \cup\left(a,b\operatorname{not\;in\;the\;same\;row}\right)\\
            0 & a,b\in\left\{ 4,5,9,10\right\}\\    0&a\left(b\right)\in\left\{ 1,2,3,6,7,8,11,12,13\right\} \\&\cup b\left(a\right)\in\left\{ 4,5,9,10\right\} 
            \end{cases};
    \end{equation}
    
    \item $i=j$:
    \begin{equation}
    \varepsilon_{ij}\left(O=X_{a}X_{b}\right)\propto\begin{cases}
    \frac{1}{8}\pi^{2}\kappa^{2} & a,b\in\left\{ 1,2,3,11,12,13\right\} \\
    \frac{1}{2}\pi^{2}\kappa^{2} & a,b\in\left\{ 6,7,8\right\} \\
    \frac{1}{4}\pi^{2}\kappa^{2} & a\left(b\right)\in\left\{ 1,2,3,11,12,13\right\} \\
     & \cup b\left(a\right)\in\left\{ 6,7,8\right\} \\
     & \cup\left(a,b\operatorname{not\;in\;the\;same\;column}\right)\\
    -\frac{19}{4}\pi^{2}\kappa^{2} & \left(a,b\right)\in\left\{ \left(1,6\right),\left(1,4\right),\left(3,5\right),\left(3,8\right),\left(11,9\right),\left(11,6\right),\left(13,8\right),\left(13,10\right)\right\} \\
    \frac{1}{4}\pi^{2}\kappa^{2} & \left(a,b\right)\in\{\left(1,5\right),\left(1,9\right),\left(1,10\right),\left(2,7\right),\left(2,9\right),\left(2,10\right),\left(3,4\right),\left(3,9\right),\left(3,10\right),\\
     & \qquad\left(11,4\right),\left(11,5\right),\left(11,10\right),\left(12,4\right),\left(12,5\right),\left(12,7\right),\left(13,4\right),\left(13,5\right),\left(13,9\right)\}\\
    \frac{3}{4}\pi^{2}\kappa^{2} & a,b\in\left\{ 4,5,9,10\right\} \\
     & \cup\left(a,b\operatorname{in\;the\;same\;column}\right)\\
    \frac{1}{2}\pi^{2}\kappa^{2} & a,b\in\left\{ 4,5,9,10\right\} \\
    & \cup\left(a,b\operatorname{not\;in\;the\;same\;column}\right)\\
    \frac{3}{4}\pi^{2}\kappa^{2} & \left(a,b\right)\in\left\{ \left(2,4\right),\left(2,5\right),\left(12,9\right),\left(12,10\right),\left(7,4\right),\left(7,5\right),\left(7,9\right),\left(7,10\right)\right\} \\
    -2\pi^{2}\kappa^{2} & \left(a,b\right)\in\left\{ \left(6,4\right),\left(6,9\right),\left(8,5\right),\left(8,10\right)\right\} \\
    \frac{1}{2}\pi^{2}\kappa^{2} & \left(a,b\right)\in\left\{ \left(6,5\right),\left(6,10\right),\left(8,4\right),\left(8,9\right)\right\} 
    \end{cases}.
    \end{equation}
\end{itemize}        

\end{widetext} 
\end{appendix}


\bibliography{DICE_Ref}
\bibliographystyle{apsrev4-1}



\end{document}